*The Performance of CRTNT Fluorescence Light Detector for Sub-EeV Cosmic Ray Observation*

Y.Bai, G.Xiao and Z.Cao
Institute of High Energy Physics 19B Yuquan Lu, Shijingshan, Beijing, 100049, P. R. China

**Abstract**

Cosmic Ray Tau Neutrino Telescopes (CRTNT) using for sub-EeV cosmic ray measurement is discussed. Performances of a stereoscope configuration with a tower of those telescopes plus two side-triggers are studied. This is done by using a detailed detector simulation driven by Corsika. Detector aperture as a function of shower energy above $10^{17}$ eV is calculated. Event rate of about 20k per year for the second knee measurement is estimated. Event rate for cross calibration with detectors working on higher energy range is also estimated. Different configurations of the detectors are tried for optimization.

## 1. Introduction

Cosmic rays observed in the energy range of $10^{16}$ eV to $10^{20}$ eV behave that their sources may switch from inside our galaxy to the larger space range[1]. The detailed modeling of the acceleration of the cosmic rays and their transportation through the space between the sources and the earth strongly depends on an accurate observation of spectrum and composition of the cosmic rays. The existing measurements have rather large discrepancies between each others[2] mainly due to a lack of a common calibration for all experiments and limited dynamic range of a single experiment. A way to overcome the difficulty is to put several independent experiments together and dedicate each of them to cover a suitable energy range and maintains good overlaps between the experiments. Using the cosmic ray events falling in the overlaps, one can cross-calibrate the detectors and achieve a complete and self-consistent measurement of the cosmic ray energy spectrum and their composition above $10^{16}$ eV. The CRTNT experiment is designed to cover the energy range from $10^{17}$ eV to $5 \times 10^{18}$ eV and co-sites with the TALE ($10^{18}$ eV~$5 \times 10^{19}$ eV)[3] and the TA (above $10^{19}$ eV)[4]. Ultimately achieve the goal.

## 2. Detector

The proposed CRTNT project uses fluorescence light telescopes analogous to the detectors of the HiRes experiment [5]. The telescopes are distributed in three groups located at three vertices of an isosceles triangle which has an eight km base line and three km height, named as FD1, FD2 and FD3, as shown in Fig. 1. At the central site FD1, twelve telescopes are used to form a "tower of power" detector that observes an area covered by 64º in azimuth and 42º in elevation starting from 3º. Those are located at the bottom vertices both have two telescopes watching into each other and over an area about 32º in azimuth and 14º in elevation starting from 44º. The total field of view covers about 58º in elevation. By comparing with two different configurations, the distances between groups are determined by simulation results discussed at the end of this paper.

A 5.0m$^2$ light collecting mirror with a reflectivity of 82% is used for each telescope. A focal plane camera is made of 16x16 pixels. Each pixel is a 40mm hexagonal photomultiplier tube that has about a 1ºx1º field of view. Each tube is read out by a 50 MHz flash ADC electronics system to measure the waveform of the shower signals. A pulse finding algorithm is developed for providing an individual channel trigger using a field programmable gate array (FPGA). The first level trigger is set by requiring the signal–noise ratio to be greater than 3.5σ, where the σ is the standard deviation of the total photo-electron noise within a running window of 320ns. The second level trigger requires at least five channels triggered within a 5x 5 running box over a single telescope camera of 16x16 pixels. The trigger condition for an event is that at least one telescope is triggered. All triggers are formed by FPGA chips. Event data from all channels are scanned from the FPGA buffers into a local Linux box.

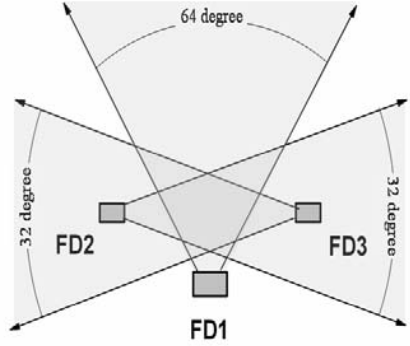

Fig.1. Configuration 344: FD1(0,0,0) FD2(-4,3,0) FD3(4,3,0)
Configuration 455: FD1(0,0,0) FD2(-5,4,0) FD3(5,4,0)

A Monte Carlo simulation program for the CRTNT detector is developed as described in the next section. A triggered event shown in the Fig. 2 is an example that has an angular track length about 50º and covers both rising and decaying stages of the shower development.

## 3. Monte Carlo Simulation

In the simulation, the incident cosmic rays are coming from all directions above the ground. The flux of cosmic rays is assumed to be isotropic and uniform in the field of view of the detector. The impact parameter, $R_p$, to the central position between the two side-trigger detectors is limited to be less than 10 km. A pure proton primary composition is assumed in the simulation. The lowest energy is set to be $2 \times 10^{16}$ eV and a $1/E$ spectrum is assumed for the detector aperture estimation and a $1/E^3$ spectrum is assumed for the resolution study.

### 3.1. Air shower simulation

Corsika 6.0 [6] is used to generate air showers in the atmosphere. A four-seasonal atmospheric model is used to describe the air density as a function of height. A big set of the simulated showers above $10^{16}$ eV are parameterized by using three parameters, shower maximum location $X_{max}$, the maximum number of shower charged particles $N_{max}$ and a dimensionless width of the shower longitudinal development function $\sigma_s$, including their energy dependence and the correlations between them. Each shower longitudinal development then is described by a function[7]

$$N_{ch}(x) = N_{max} \exp\{-\frac{2(x-x_{max})^2}{\sigma_s^2 (2x_{max}+x)^2}\} \quad,$$

where $N_{ch}(x)$ is the number of charged shower particles at the slant atmospheric depth x.

### 3.2. Photon production and light propagation

Charged shower particles excite the nitrogen molecules as they pass through the atmosphere. The deexcitation of the molecules generates ultra-violet fluorescence light. The number of fluorescence photons is proportional to the shower size, and these photons are emitted isotropically. The shower simulation carried out in this paper assumes a fluorescence light spectrum according to a recent summary of world-wide measurements [8], including the dependence of the yield on the atmospheric pressure and temperature. Since energies of charged shower particles are higher than critical energy, shower particles generate Cerenkov photons at every stage of the shower development. The accumulated Cerenkov light is concentrated in a small forward cone, therefore intensity of the light is much stronger than the fluorescence light along the shower direction. A significant part of the Cerenkov light can be scattered out via Rayleigh and Mie scattering during the whole shower development history. The fraction of this light scattered in the direction of the detector can also make a noticeable contribution to detector triggering. Cerenkov light generation and scattering is fully simulated. A detailed description of the calculation can be found in [1] and references therein.

Shower charged particles and therefore fluorescence light photons, spread out laterally following the NKG distribution function. The Molier unit of the NKG function is about 95 m at about 1500 m a.s.l. Photons originating from Cerenkov radiation have an exponential lateral distribution from the axis of the shower. Therefore, photons coming from a shower are spread over a range of directions around the shower location in the sky due to its longitudinal motion and lateral extension. A ray tracing procedure is carried out to follow each photon to the PMT's from the photon source location. All detector responses are considered in the ray tracing procedure, including mirror reflectivity, UV filter transmission, quantum efficiency of photo-cathode, location-sensitive response function of the photo-cathode and optical effects associated with the off-axial and defocusing effects. Sky noise photons, 40ph/μsec/m$^2$, are randomly added in this ray tracing procedure both in time and arrival directions. The uncertainty associated with the varying weather conditions is negligible for the Rayleigh scattering. Scattering due to aerosols is more dependent on weather conditions. However, for a detector that has an aperture within 6 km, the aerosol scattering contribution to light extinction is close to be its minimum. The uncertainty in the triggering efficiency due to weather conditions is thus small. In the simulation, an average model [9] of aerosol scattering for western US deserts is employed.

## 4. Detector Aperture and Event Rate

The detector aperture is estimated using more than 10$^4$ simulated events. The triggering condition is the signal to noise ratio being greater than 3.5 for the tower detector and 2.5 for the side-trigger detectors. The trigger aperture is shown in the Fig. 3 as the dash line. It is noticed in

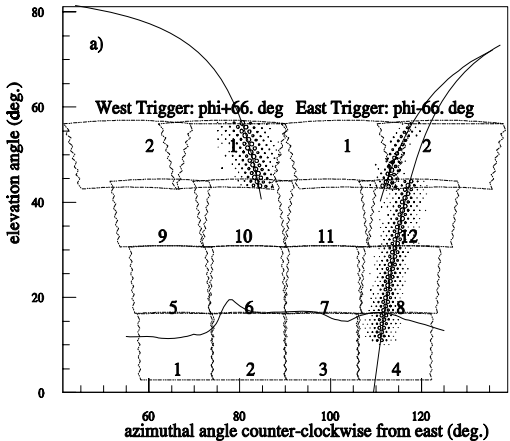 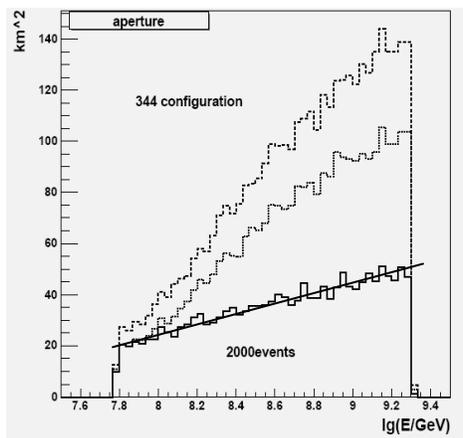

Fig. 2. An example event triggered by all three CRTNT detectors

Fig. 3. CRTNT detector apertures, see text for details.

Fig. 2 that the tower detector has two vertical edges that could cause very biased measurement of the triggered showers. A cut on those events that fall in the detector edges is applied for good shower reconstruction quality. We also made cuts on the Cerenkov light dominant events by removing tubes that have viewing angles less than 20º and requiring the track lengths being still greater than 6º. The other cut is that the amplitude weighted average vector of the fired tubes in the side-trigger detectors must be in the neighborhood (<0.5º) of a true shower-detector-plane. The last cut guarantees the shower geometric reconstruction resolution to be better than 0.7º. After all the cuts for the quality of shower measurement, the aperture is calculated shown in Fig. 3 as the dot line. For energy above $10^{18}$eV, the aperture is greater than 100 km²sr. This allows that there are about 700 common events per year to be observed by both CRTNT and TALE for cross calibration between the two detectors. Based on this, we have 30k events per year measured above $10^{17}$eV.

In order to measure the structure of the cosmic ray energy spectrum around $10^{17.5}$eV, an aperture independent to the energy is ideal. A geometric constraint, i.e. $R_p$<6km, makes a significant suppression in the high energy range and remains the low energy aperture unchanged, as the solid lines shown in Fig. 3, where the straight line is a fit to the histogram for eye guiding only. With such a flat aperture, about 20k reconstructible events per year are expected above $10^{17}$eV.

## 5. Comparison between different configurations

Comparing with other experiment, the CRTNT detector has an advantage of mobility. This allow us easily change the detector configuration for different physics. Two configurations of the CRTNT detector are compared for optimizing the aperture of the detector. In the simulation described above, the detector is set symmetrically at three sites as shown in Fig. 1. As described in the beginning of this paper, the distances from the detector sites to the bisection of the connection between FD2 and FD3 are 3 km, 4 km and 4 km, respectively. This configuration is denoted as 344 in Fig. 4. The other configuration investigated is to put all three detectors by 1 km further apart, i.e. the distances become 4 km, 5 km and 5 km, and denoted as 455. The simulation is repeated for the configuration 455. Results suggest that the aperture for configuration 455 is larger than configuration 344 at higher energy while the aperture remains no change at low energies around $10^{17}$ eV. This feature provides more events above $10^{18}$ eV (about 15%) for cross-calibration with other detectors, such as TALE.

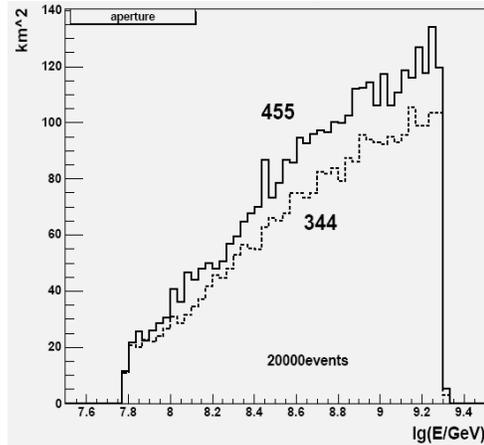

Fig. 4. Apertures of two configurations after quality cuts.

## 6. Acknowledgment
This work is supported by Knowledge Innovation fund (U-526) of IHEP, China and by Hundred